# Phase-Modulated Degenerate Parametric Amplification Microscopy

Yunan Gao,[1,†] Aaron J. Goodman,[2] Pin-Chun Shen,[3] Jing Kong,[3] and William A. Tisdale[1,*]

1 Department of Chemical Engineering, 2 Department of Chemistry, 3 Department of Electrical Engineering and Computer Sciences, Massachusetts Institute of Technology, Cambridge, Massachusetts 02139, USA

E-mail: tisdale@mit.edu

Phone: +1.617.253.4975 | Fax: +1.617.258.5766

**Abstract**

Second-order nonlinear optical interactions, including second harmonic generation (SHG) and sum-frequency generation (SFG), can reveal a wealth of information about chemical, electronic, and vibrational dynamics at the nanoscale. Here, we demonstrate a powerful and flexible new approach, called phase-modulated degenerate parametric amplification (DPA). The technique, which allows for facile retrieval of both the amplitude and phase of the second-order nonlinear optical response, has many advantages over conventional or heterodyne-detected SHG, including the flexibility to detect the signal at either the second harmonic or fundamental field wavelength. We demonstrate the capabilities of this approach by imaging multi-grain flakes of single-layer $MoS_2$. We identify the absolute crystal orientation of each $MoS_2$ domain and resolve grain boundaries with high signal contrast and sub-diffraction-limited spatial resolution. This robust all-optical method can be used to characterize structure and dynamics in organic and inorganic systems, including biological tissue, soft materials, and metal and semiconductor nanostructures, and is particularly well-suited for imaging in media that are absorptive or highly scattering to visible and ultraviolet light.

Second harmonic generation (SHG) and sum-frequency generation (SFG) microscopy are powerful tools for characterizing structure and dynamics in low-dimensional materials, biological tissues, and at interfaces between centrosymmetric media.[1-6] However, conventional SHG imaging reveals only the squared magnitude of the second harmonic field – not its phase.[1-5] To overcome this limitation, phase-sensitive variants of SHG and SFG microscopy, based on the principle of heterodyne detection, have been developed;[7-13] examples include interferometric SHG (I-SHG) microscopy[7-10] and second-harmonic digital holographic microscopy.[10, 13]

In heterodyne SHG, a second harmonic signal field is generated within the sample and then mixed with a reference field to retrieve its amplitude and phase. In our previous work,[14] we showed that when an optical field and its second harmonic are temporally coincident within the sample – as is sometimes the case in I-SHG[8-11, 15, 16] – the interaction can also be described as stimulated SHG or parametric amplification. Significantly, we showed that the direction of power flow between the two input fields can be changed by varying the relative optical phase.

Here, we take advantage of the phase sensitivity of second-order nonlinear optical interactions by introducing a high-frequency phase modulation scheme, which produces an intensity modulation in the output beams that can be detected using a standard silicon photodiode and lock-in amplifier. The resulting technique, which we refer to as phase-modulated degenerate parametric amplification (DPA), can retrieve both the magnitude and phase of the second-order nonlinear optical response. The method can be implemented under ambient lighting conditions, is compatible with high-speed laser scanning systems, and provides the enabling option to detect the signal at either the fundamental or second harmonic frequency. Moreover, the use of 100 fs duration laser pulses renders the technique amenable to spatially resolved ultrafast spectroscopy.[1, 17-19]

We demonstrate the utility of DPA microscopy by imaging multi-grain flakes of single-layer MoS2 grown by chemical vapor deposition (CVD).[20] We successfully identify the absolute crystal orientation of each MoS2 domain and resolve grain boundaries with high signal contrast and sub-diffraction limited spatial resolution – information that is usually obtainable only by high-resolution electron microscopy.

**Theoretical Foundation**

The interaction of a fundamental field with its second harmonic in a nonlinear medium was first described by Armstrong *et al.* in 1962.[21] The two fields exchange energy according to the coupled equations,

$$\frac{du_\omega}{d\varsigma} = u_\omega u_{2\omega} \sin(\varphi)$$
$$\frac{du_{2\omega}}{d\varsigma} = -u_\omega^2 \sin(\varphi)$$
(1)

where $u_\omega$ and $u_{2\omega}$ are the normalized field amplitudes of the fundamental and the second harmonic fields, $\varsigma$ is a normalized propagation distance that depends on the second-order nonlinear susceptibility $\chi^{(2)}$, and $\varphi = 2\phi_\omega - \phi_{2\omega}$ is a defined relative optical phase.[21, 22]

When only the fundamental field is incident on the sample, spontaneous SHG occurs. When $u_{2\omega}$ is non-zero, the incident second harmonic field stimulates either second harmonic generation (SHG) or difference frequency generation (DFG), depending on the relative optical phase of the fundamental and second harmonic fields.[14] As a result, the incident fundamental field is either amplified or attenuated by power transfer from (to) the second harmonic field as the relative optical phase is cycled. This interaction is known as degenerate parametric amplification – a specific case of optical parametric amplification (OPA) wherein the signal and idler frequencies are degenerate.[22] As a parametric process, no energy is exchanged with the sample itself, and either beam can be detected to retrieve information about the sample.

The salient difference between DPA and heterodyne-detected SHG is the location of the optical interaction. Classically, heterodyne detection (or second harmonic interferometry)[7-10] is described as optical interference at the *detector*, whereas parametric amplification (or stimulated SHG)[14] is described as nonlinear wave mixing within the *sample*. As a result, DPA leads to depletion or amplification of the residual fundamental field, whereas heterodyne SHG does not.[23]

**Phase-Modulated Degenerate Parametric Amplification Microscopy**

The phase-modulated degenerate parametric amplification (DPA) microscopy technique, as illustrated schematically in Fig. 1, is performed by rapidly oscillating the relative optical phase between co-incident $\omega$ and $2\omega$ fields so that power transfer changes direction at MHz frequency. The transmitted (or reflected) laser power of the fundamental ($\lambda = 830$ nm) or second harmonic ($\lambda = 415$ nm) beam is detected using a standard photodiode, and the magnitude and phase of the DPA signal is retrieved using a lock-in amplifier.

We found that high-frequency phase modulation overcomes difficulties associated with low-frequency phase noise from vibrations, air currents, and other instabilities on the optical table, enabling phase-sensitive measurements to be performed in a laboratory environment that did not have sufficient passive phase stability for heterodyne-SHG. Furthermore, careful choice of modulation depth (exactly $2\pi$) and waveform (saw tooth) is essential for making the measurement robust against long-term thermal drift (see Supplementary Note 1). Phase modulation also overcomes many challenges that are normally associated with high-frequency *intensity* modulation, such as fluorescence, transient heating, and differential transmittance/reflectance (*i.e.* third-order pump-probe signals).

The excellent signal contrast achievable with DPA microscopy is demonstrated on a single-layer $MoS_2$ flake, shown in Fig. 1c-d. For these measurements we used cross-polarized fundamental and second harmonic laser pulses (~110 fs) generated by a Ti:sapphire laser and detected the residual fundamental field intensity in transmission mode using a silicon photodiode.

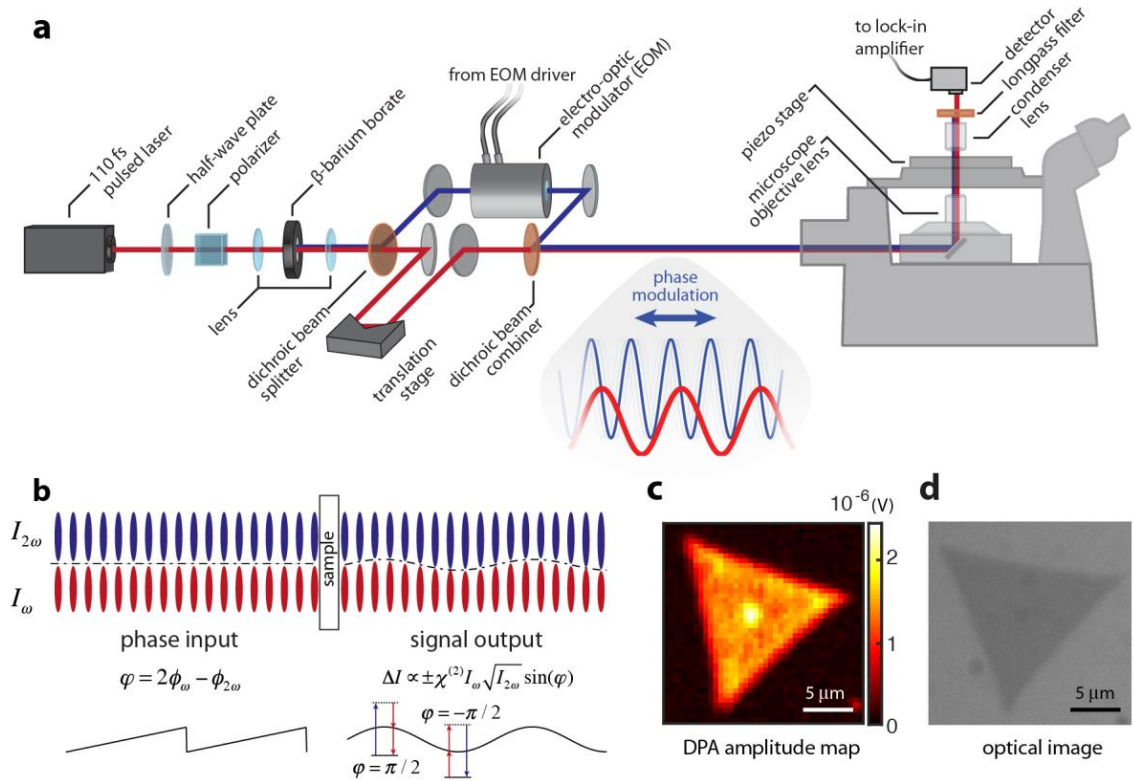

**Figure 1. Phase-modulated degenerate parametric amplification (DPA) microscopy.** (a) Schematic illustration of the optical apparatus and associated components. In this embodiment, λ = 830 nm and 415 nm pulses generated by a Ti:sapphire laser are cross-polarized at the sample and the fundamental field intensity is measured in transmission using a standard photodiode. (b) Illustration of the working principle of the technique: the electro-optic modulator (EOM) modulates the relative optical phase, $\varphi$, between the two laser pulses at 1 MHz according to a saw tooth waveform at $2\pi$ modulation depth, generating a sinusoidal variation of the intensity in each field. (c) DPA signal amplitude map of a triangular monolayer $MoS_2$ flake. (d) Bright-field optical image of the same flake.

**Phase-Sensitive Imaging of MoS$_2$**

One of the most critical challenges in characterizing two-dimensional materials is unambiguous determination of the absolute crystal orientation.[24] This information is essential for the fabrication of anisotropic 2D heterostructures and devices,[25] and for the understanding of 2D materials growth processes.[26] In addition, grain boundaries themselves provide opportunities for the discovery of novel physical phenomena.[27-29] Photoluminescence and Raman are not strongly sensitive to domain rotations, and polarized SHG cannot distinguish between domains with mirror symmetry.[1, 27, 30-32] Advanced transmission electron microscopy (TEM) techniques such as high resolution TEM combined with electron diffraction or scanning-TEM (STEM) can be used to precisely determine crystal orientation and identify grain boundaries, but these techniques are not compatible with optical and electronic substrates,[33-35] thus preventing *in-situ* examination and requiring sample preparations that are delicate and tedious. Here, we show that DPA microscopy is a facile way to retrieve this structural information.

Monolayer MoS$_2$ has $D_{3h}$ symmetry with non-vanishing second-order nonlinear susceptibility tensor components of $\chi^{(2)}_{yyy} = -\chi^{(2)}_{yxx} = -\chi^{(2)}_{xxy} = -\chi^{(2)}_{xyx} = \chi^{(2)}$,[1, 36] where the coordinate system is defined with respect to the MoS$_2$ atomic lattice as shown in Fig. S5. For the case of cross-polarized fundamental and second harmonic fields, the change in fundamental field intensity is given by (see Supplementary Note 2),

$$\Delta I_\omega \propto \chi^{(2)} \cos(3\theta) I_\omega \sqrt{I_{2\omega}} \sin(2\phi_\omega - \phi_{2\omega}) \qquad (2)$$

where $\theta$ denotes the orientation of the MoS$_2$ crystal with respect to the fundamental field polarization, as illustrated in Fig. 2, and $\Delta I_{2\omega} = -\Delta I_\omega$. Note that both the sign of $\chi^{(2)}$ and the orientation of the MoS$_2$ crystal determine whether $\Delta I_\omega$ is positive or negative. A 180º rotation of the crystal ($\theta = 0°$ and $180°$) will generate an equal amplitude signal, but opposite lock-in phase, as demonstrated in Fig. 2.

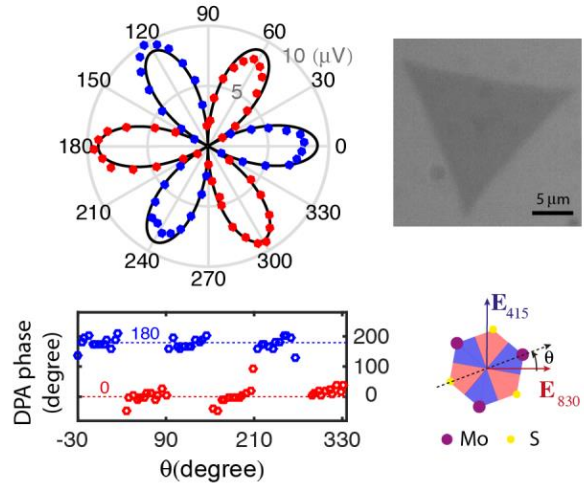

**Figure 2. Effect of MoS$_2$ crystal orientation on DPA amplitude and phase.** Rotational dependence of the DPA amplitude (upper panel) and phase (lower panel) for the triangular MoS$_2$ single-crystal flake shown in the bright field image. Red and blue colors in two panels indicate a lock-in phase of 0° or 180°, respectively.

In lower-symmetry materials, DPA can be used to determine each component of the full nonlinear susceptibility tensor. Each combination of wave vector (with respect to a defined crystal axis), fundamental field polarization, and second harmonic polarization can be chosen to isolate a specific $\chi^{(2)}$ matrix element.[14] By systematically varying the crystal orientation and input field polarizations, the complete second-order nonlinear susceptibility tensor can be mapped. This approach can be extended to surface and interfacial phenomena in centrosymmetric media, too.

In Fig. 3, we characterize a CVD-grown monolayer MoS$_2$ flake composed of multiple crystal grains with varied orientation. The DPA amplitude and phase maps are shown in Fig. 3a and b. The combined amplitude and phase information was used to determine the absolute orientation of each crystal grain relative to the polarization of the incident fundamental field, as shown in Fig. 3c (see Supplementary Note 3). After identifying the crystal orientation of all grains, the tilt angle between adjacent grains could be determined. The grains I, II, III and IV are commonly observed neighboring mirror twins with a relative tilt angle of 60°, and their grain boundaries are comprised of 4|8-member or 4|4-member rings.[33, 35] The grains V, VI and VII have a relative tilt angle of 51°; IV and V have a relative tilt angle of 27°; and I and VII have a relative tilt angle of 33°. These grain boundaries have been predicated by atomistic simulations or observed by other techniques, but the atomic structures have not been identified.[27, 29, 30]

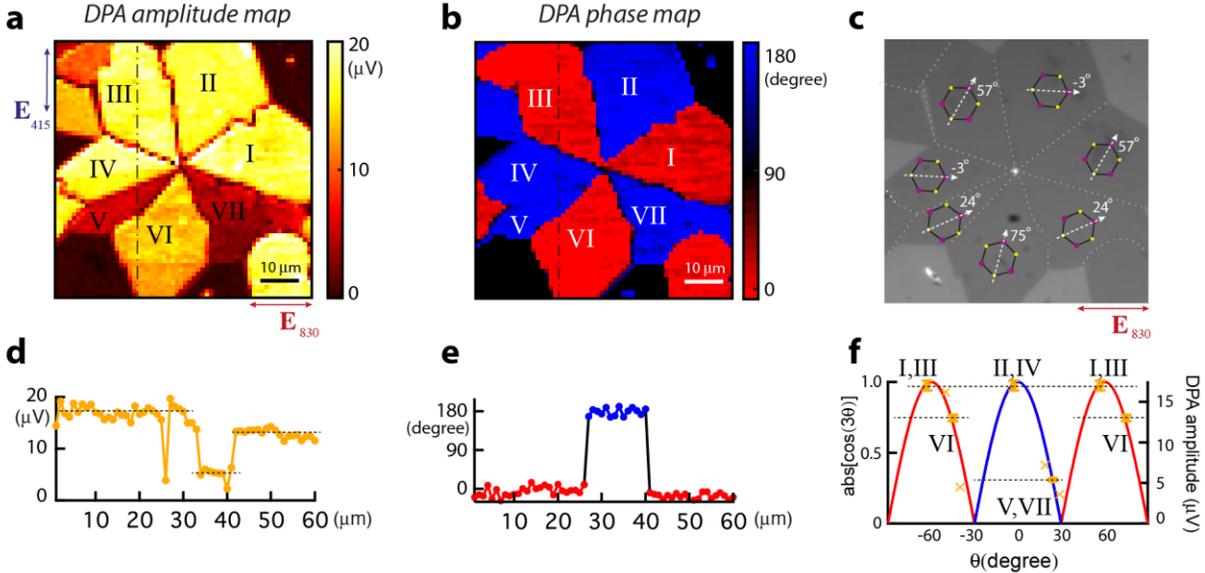

**Figure 3. Mapping grain orientations in MoS$_2$ crystals.** (a,b) DPA amplitude and phase maps of a multi-grain MoS$_2$ flake. (c) Bright-field image of the flake mapped in panels a and b, with the boundaries and the absolute orientation of each crystal grain, relative to the polarization of the fundamental field ($E_{830}$), labeled. (d,e) Line cuts taken along the dash-dotted lines through the images in panels a and b. (f) Universal curve used to determine the crystal orientation of each grain labeled in panel c.

**DPA imaging of grain boundaries**

We demonstrate the capability of resolving grain boundaries with high signal contrast and spatial resolution using DPA microscopy. In Fig. 4, we show DPA amplitude and phase maps in the vicinity of a mirror twin boundary in a monolayer MoS$_2$ flake. The magnitude of the DPA signal scales as $S \propto \chi^{(2)} I_\omega \sqrt{I_{2\omega}}$, according to its dependence on the fundamental and second harmonic field intensities (see Supplementary Note 1). However, the DPA amplitude goes to zero at the grain boundary due to destructive interference between the second harmonic fields originating from the left and right sides of the boundary. DPA microscopy has superior signal contrast compared to conventional SHG, which can also be used to resolve the boundary based on a similar mechanism (see Supplementary Note 5).[1]

Line scans of the amplitude (Fig. 4c) and phase (Fig. 4f) reveal diffraction-limited resolution in the amplitude map (see Supplementary Note 4) and sub-diffraction limited resolution in the phase map. The excellent spatial resolution of the phase map is due to the binary nature of this signal; it must be either 180º or 0º, and the spatial resolution is ultimately limited by the vanishing signal intensity at the boundary.

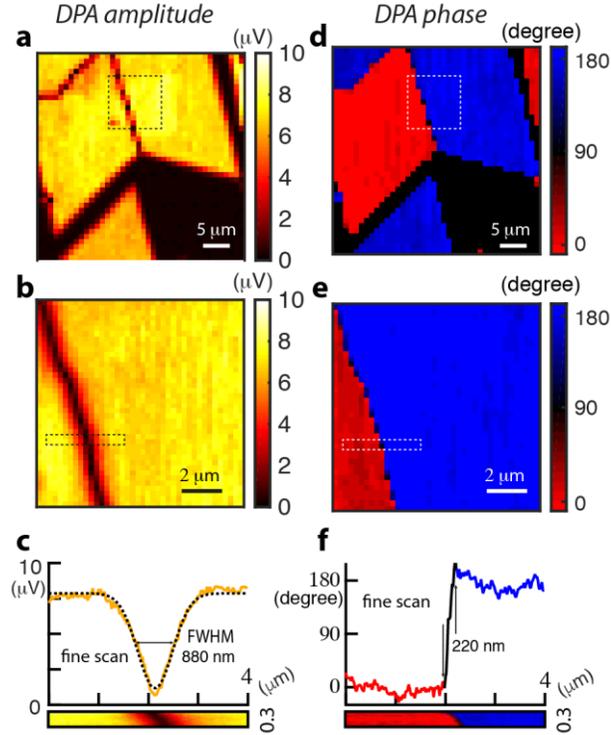

**Figure 4. Characterization of grain boundaries.** DPA amplitude (a-c) and phase (d-f) mapping near a mirror twin boundary in a monolayer $MoS_2$ flake.

**Comparison between DPA and SHG**

In Fig. 5, we compare DPA imaging to conventional SHG microscopy. In all images, a constant acquisition time of 1 second per pixel was used. The SHG implementation used for comparison is shot-noise-limited, such that the background SHG signal arises only from dark counts of a photomultiplier tube (Hamamatsu R4220P). In contrast, the DPA signal represents a small fluctuation on top of a large background of incident $\omega$ or $2\omega$ light that is captured by the photodiode. Because the DPA signal intensity scales with incident $2\omega$ laser power at the same rate as the noise power contained in the incident field, the signal-to-noise ratio for DPA imaging is theoretically equivalent to heterodyne SHG. As with heterodyne SHG, the DPA signal-to-noise ratio can exceed that of conventional SHG in experiments with high signal background levels.[37]

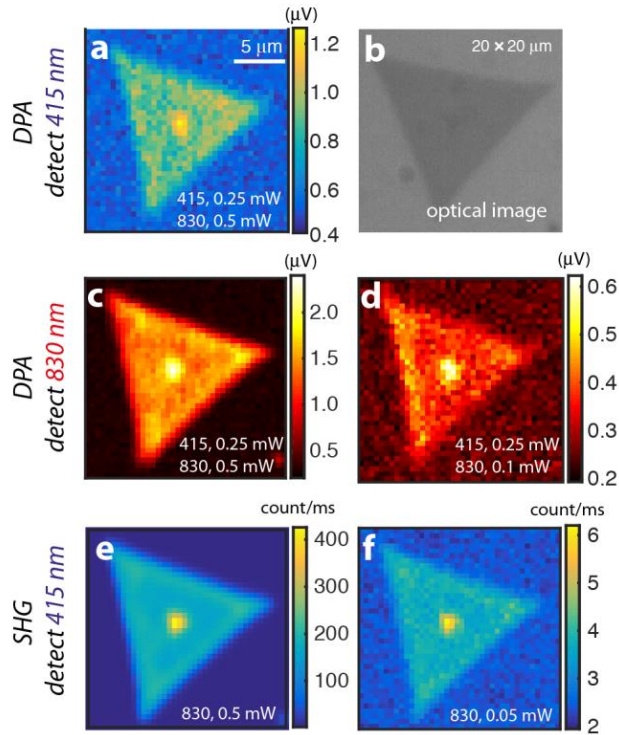

**Figure 5. Sensitivity comparison between DPA and conventional SHG.** (a) DPA amplitude map acquired with detection at the second harmonic field wavelength, $\lambda = 415$ nm. (b) Bright-field optical image of the same $MoS_2$ flake. (c, d) DPA amplitude maps acquired with detection at the fundamental field wavelength, $\lambda = 830$ nm, at two different laser powers. (e, f) Conventional SHG intensity maps acquired using different laser power.

DPA offers additional advantages over polarized SHG or heterodyne SHG. First, the signal is contained in both the fundamental and harmonic fields, so that either beam can be detected as experimental circumstances necessitate. This is particularly useful if the sample, solvent, or substrate is highly scattering or absorptive at the second harmonic wavelength. Furthermore, the signal intensities are large enough to be detected with a standard silicon photodiode under normal room lighting, and the high-frequency phase modulation scheme enables phase-sensitive measurements to be performed in laboratory environments that do not have sufficient passive phase stability for interferometry. Finally, high-frequency (MHz) phase modulation and lock-in detection is compatible with high-speed laser scanning microscopes. Wide-field signal acquisition could also be achieved (without raster scanning the laser spot) by using microlens arrays with chip-integrated lock-in pixels or tuned amplifier arrays.[38, 39] Beyond 2D materials characterization, we expect DPA to find use in bioimaging, interfacial spectroscopy, and ultrafast microscopy.[3-5, 17, 18]

**Materials and Methods**

*Phase-modulated degenerate parametric amplification (DPA) microscopy.* See Supplementary Note 6 for detailed optical layout. A Ti:sapphire oscillator (Coherent Mira HP) generated 830 nm laser pulses of ~110 fs pulse duration at 76 MHz repetition rate. The $\lambda = 830$ nm laser beam passed through a small aperture (to produce a near-Gaussian beam), a half-wave plate, a Glan-Taylor polarizer (GT5, Thorlabs), then was focused into a 0.1 mm thick $b$-barium borate crystal (BBO, Type 1, Eksma Optics) by a 50 mm focal length lens (~8 nJ pulse energy) to generate $\lambda = 415$ nm laser pulses (~40 pJ pulse energy). The output beams were re-collimated by a 50 mm focal length lens, and then separated by a dichroic beam splitter (042-0845，>99.5% reflectivity @ 800nm，Eksma Optics). Residual 830 nm light was removed from the 415 nm beam using a colored glass filter (FGS550, Thorlabs). The 415 nm or the 830 nm beam was sent to the EOM, depending on which field was used for detection (see details below and Fig. S11); the detected beam bypassed the EOM. One of the two beams was sent to an optical delay line consisting of mirrors mounted on a linear translation stage (462-X-M, Newport Corporation) driven by a piezoelectric inertia actuator (ZBT225B, Thorlabs). The second harmonic and fundamental paths were re-combined by a beam combiner (042-4805 Eksma Optics, >99.5% reflectivity @ 390-410 nm), and were directed into an inverted optical microscope (Nikon Ti-U).

The two beams were focused onto the sample by a microscope objective lens (Nikon, CFI S Plan Fluor ELWD, 40×, 0.6 numerical aperture), and then directed to an amplified Si photo-detector (PDA 36A, Thorlabs) by a 25 mm focus length condenser lens. When the 830 nm (415 nm) laser beam was detected, a long (short) pass filter FGL550S, Thorlabs (FGB 39, Thorlabs) was place in front of the detector. The signal from the photodetector was fed directly into a lock-in amplifier (HF2LI, Zurich Instruments), and the output was recorded and analyzed by a computer. For spatial mapping, the sample was scanned relative to the stationary focal point using a piezo stage (P-545.xR8S PI nano XY Piezo System, Physikinstrumente). For the sample rotation dependent experiments, a motorized precision rotation stage (PRM1Z8, Thorlabs) was used.

A function generator (4063, BK Precision) generated the saw-tooth wave for the EOM driver (Model 275, Conoptics) and supplied the reference signal for the lock-in amplifier. The EOM was a potassium dideuterium phosphate (KD*P) phase modulator (M350-160 phase, Conoptics). To minimize phase noise, the separated beam paths (dual beam paths from separator to combiner, including the EOM) were placed inside a homemade box to minimize the effects of external disturbances. Inside the box, the EOM was placed on a water-cooled breadboard (Thorlabs) fed by a re-circulating chiller (Coherent).

*MoS$_2$ monolayer chemical vapor deposition (CVD).*[20] The growth substrate was a 285nm SiO$_2$/Si wafer, cleaned by deionized water, acetone, and isopropyl alcohol sequentially before growth. Perylene-3,4,9,10-tetracarboxylic acid tetrapotassium salt (PTAS) molecules were used as the seeding promoter and two additional clean SiO$_2$/Si substrates were coated with PTAS as seed reservoirs. The growth substrate was then suspended between those two PTAS seed reservoirs. All of these substrates were faced down and placed on a crucible containing a molybdenum oxide (MoO$_3$, 99.98%) precursor. This MoO$_3$ precursor was put in the middle of a 1 inch quartz tube reaction chamber and another sulfur powder (99.98%) precursor was placed upstream, 14 cm away from MoO$_3$ precursor, in the quartz tube. Before heating, the CVD system was purged using 1000 sccm of Ar (99.999% purity) for 5 min, and then 20 sccm of Ar was flowed into the system as a carrier gas. Next, the temperature of the reaction chamber was increased to 625 °C at a rate of 30 °C min$^{-1}$. The monolayer MoS$_2$ was synthesized at 625 °C for 3 min under atmospheric pressure. The temperature at the position where the sulfur was located was ~ 180 °C during growth. Finally, the system was cooled down to room temperature quickly using an electric fan. During the cooling process, 1000 sccm Ar flow was introduced to remove the reactants, preventing further unintentional reactions.

*MoS$_2$ transfer process.* The CVD-grown monolayer MoS$_2$ was transferred onto transparent glass for all the optical measurements by a wet transfer process. First, poly-methylmethacrylate (950 PMMA A4) was spin-coated (4000 rpm for 1 min) onto the as-grown monolayer MoS$_2$ samples. Next, the PMMA/MoS$_2$/SiO$_2$/Si stack was placed in an aqueous KOH solution, and the solution was then heated up to 85 °C. After the SiO$_2$ layer was etched away, the PMMA/MoS$_2$ stack was separated from the substrate and remained floating on the solution. The PMMA/MoS$_2$ film was then placed in distilled water using a glass slide for 20 min to remove the KOH residue. This rinsing step was repeated three times. After that, the PMMA/MoS$_2$ film was transferred onto a transparent glass substrate, and was then baked at 80 °C for 10 min and 130 °C for another 10 min. This baking step can remove moisture and enhance the adhesion between MoS$_2$ and the substrate. Finally, the PMMA/MoS$_2$/glass stack was immersed in acetone for 12 hr to remove the PMMA layer.

**Corresponding Author**

* E-mail: tisdale@mit.edu

Phone: +1.617.253.4975 | Fax: +1.617.258.5766

**Present Addresses**

† State Key Laboratory for Mesoscopic Physics and School of Physics, Peking University, Beijing 100871, China.

**Author Contributions**

Y.G., A.J.G., and W.A.T. conceived of the technique and designed the experiments. Y.G. built the instrument and collected and analyzed the data. P.S. and J.K. supplied the $MoS_2$ samples and helped interpret results. All authors contributed to the writing of the manuscript.

**Acknowledgements**

We thank Richard Kocka at Conoptics Inc. for helpful discussions related to the EOM phase modulator. This work was supported by the U.S. Department of Energy, Office of Basic Energy Sciences, under Award No. DE-SC0010538. A.J.G. acknowledges partial support from the National Science Foundation under Grant No. 1122374. P.S. and J.K. acknowledge the support from the Center for Energy Efficient Electronics Science (NSF Award 0939514). Y.G. acknowledges partial support from the Office of China Postdoctoral Council under the international Postdoctoral Exchange Fellowship Program No. 20140040.

**Supporting Information**

Supporting Information available: mathematical description of the lock-in signal and its dependence on the optical phase; nonlinear wave mixing in MoS$_2$; determining the absolute MoS$_2$ crystal orientation; signal contrast comparison between DPA microscopy and conventional SHG; spatial resolution of the instrument; optical layout and electronic components.

TOC:

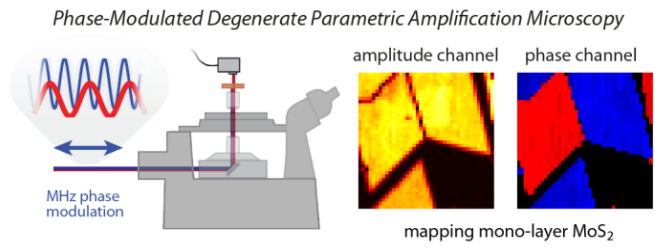

# Supplementary Information for:

# Phase-Modulated Degenerate Parametric Amplification Microscopy

*Yunan Gao[1,†], Aaron J. Goodman[2], Pin-Chun Shen[3], Jing Kong[3], and William A. Tisdale[1,*]*

1 Department of Chemical Engineering, 2 Department of Chemistry, 3 Department of Electrical Engineering and Computer Sciences, Massachusetts Institute of Technology, Cambridge, Massachusetts 02139, USA

*e-mail: tisdale@mit.edu

Contents:

**Supplementary Note 1: Mathematical description of the lock-in signal and its dependence on the optical phase**

A lock-in amplifier takes an input signal $s(t)$, multiplies it by a reference signal, and integrates over a certain time to extract a signal with the same frequency as the reference. Two-phase lock-in amplifiers can extract both the amplitude and relative phase of the input signal. For a sine wave reference integrated over a full $2\pi$ cycle, the lock-in output is:

$$Y = \frac{1}{2\pi}\int_0^{2\pi} \sin(t)s(t)dt \qquad R = \sqrt{X^2 + Y^2}$$
$$X = \frac{1}{2\pi}\int_0^{2\pi} \cos(t)s(t)dt \qquad \vartheta = \arctan\left(\frac{Y}{X}\right)\text{,}$$
(S1)

where $\sin(t)$ is the reference, $s(t)$ is the input signal, $R$ is the lock-in amplifier output amplitude, and $\vartheta$ is the lock-in amplifier output phase, which is the phase difference between the signal and reference.

For a thin (0.1 mm) BBO crystal or monolayer MoS$_2$, perfect phase matching condition applies, and the relative optical phase of the interacting fields does not depend on the propagation distance through the sample. The fundamental and harmonic fields exchange energy according to[1,2]

$$\frac{du_\omega}{d\varsigma} = u_\omega u_{2\omega}\sin\varphi,$$
$$\frac{du_{2\omega}}{d\varsigma} = -u_\omega^2\sin\varphi,$$
$$\varphi = 2\phi_\omega - \phi_{2\omega}.$$
(S2)

where $u_\omega$ and $u_{2\omega}$ are the normalized field amplitude of the fundamental and the second harmonic fields, $\varphi$ is a defined relative optical phase, and $\varsigma = z/l$ is the normalized propagation distance through the nonlinear medium. $l$ is the characteristic distance over which the two fields exchange energy, which is much longer than the sample thickness being studied. Solving Eq. S(2), we find that the signal input to the lock-in amplifier varies according to the relations,

$$s_\omega(t) = \Delta I_\omega(t) \propto u_\omega^2 u_{2\omega}\sin\varphi(t)$$
$$s_{2\omega}(t) = \Delta I_{2\omega}(t) \propto -u_\omega^2 u_{2\omega}\sin\varphi(t)\text{,}$$
(S3)

which tells us how the signal $s(t)$ depends on the functional form of the optical phase modulation, $\varphi(t)$. By inspection, we see that a linear variation of the optical phase $\varphi(t)$ will result in a sinusoidal variation of the fundamental or harmonic field intensity.

For the phase modulation scheme to be stable, the lock-in output signal must not be sensitive to slowly drifting (DC) optical phase differences between the fundamental and harmonic fields. If the signal $s(t)$ has a sine form, the lock-in output is

$$Y = \frac{1}{2\pi}\int_0^{2\pi} \sin(t)\sin(At+\Theta)dt$$
$$X = \frac{1}{2\pi}\int_0^{2\pi} \cos(t)\sin(At+\Theta)dt$$
(S4)

where $A$ is an arbitrary depth, and $\Theta$ is an arbitrary relative phase. If $A=1$,

$$Y = \frac{1}{2\pi}\int_0^{2\pi} \sin(t)\sin(t+\Theta)dt = \frac{1}{2}\cos\Theta,$$
$$X = \frac{1}{2\pi}\int_0^{2\pi} \cos(t)\sin(t+\Theta)dt = \frac{1}{2}\sin\Theta,$$
(S5)
$$R = \frac{1}{2}, \quad \vartheta = \frac{\pi}{2} - \Theta.$$

Equation S(5) indicates that in this case the lock-in amplifier amplitude output is not sensitive to the arbitrary input phase $\Theta$, which in reality could drift due to fluctuation of experimental conditions, such as temperature and some mechanical factors. If $A \neq 1$, $R$ changes with $\Theta$, as shown in Fig. S1. $A=1$ means the phase modulation $\varphi(t)$ is a linear $2\pi$ modulation, as shown in Fig. S1b. Figure S1d shows the dependence of $R$ on $A$ and $\Theta$.

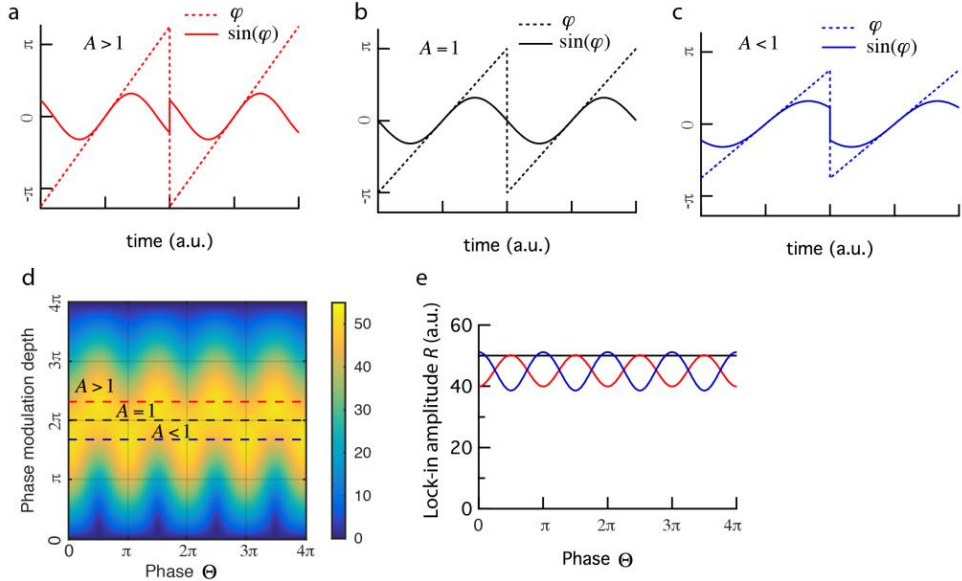

**Figure S1.** (a,b,c) Saw-tooth phase modulation forms (dashed lines) and the resulting signal modulation forms of $\sin(\varphi)$ (solid lines) for three cases of modulation depth of $A$. (d) Contour plot of the dependence of $R$ in $A$ and $\Theta$. (e) $R$ dependence in $\Theta$ along the dashed lines in panel d.

In Fig. S2, we compare simulations and experiment results for the saw tooth wave phase modulation. We introduced the phase $\Theta$ by a static bias from the EOM driver. The phase modulation depth was adjusted by the amplitude of the saw-tooth wave, which was generated by a function generator, and amplified by the EOM driver. In practice, a stable lock-in amplitude required a slightly higher modulation depth than $2\pi$, as shown in Fig. S2, due to bandwidth limitations of the EOM.

An ideal saw tooth wave is composed of an infinite number of higher-frequency sine components,

$$x(t) = \frac{C}{2} - \frac{C}{2}\sum_{k=1}^{\infty}(-1)^k \frac{\sin(2\pi kft)}{k}, \tag{S6}$$

which cannot be realized in practice due to bandwidth limitations of the EOM driver. We found that including Fourier components up to $k = 20$ in our simulations gave good agreement with the experimental results (Fig. S2). Fig. S2a shows a simulation contour, and Figs. S2b,c show comparisons between simulations and experimental results at positions along two horizontal and two vertical line cuts indicated by the dashed lines. Figure S2d presents a stability test trace over two hours, showing stability against long-term phase drift.

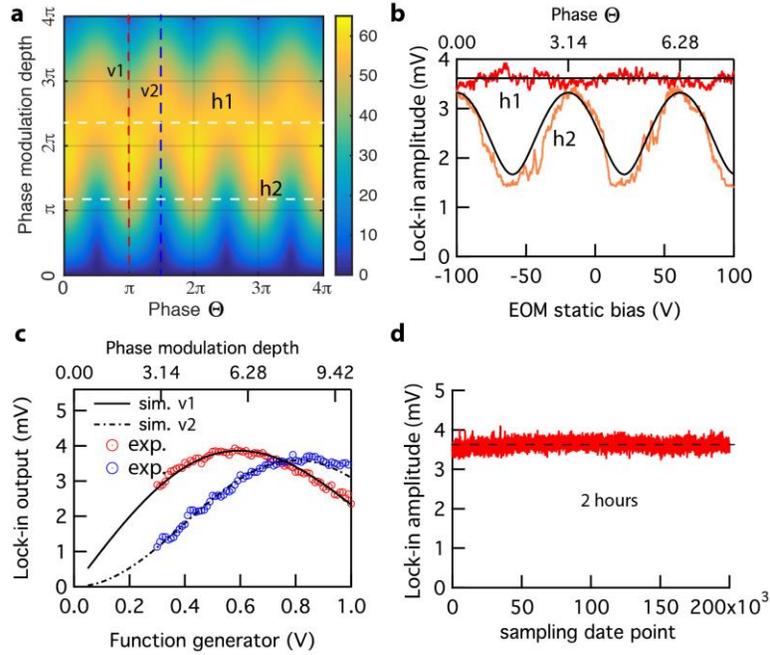

**Figure S2.** Comparison between simulations and experimental results for saw-tooth wave phase modulation. (a) Contour plot of simulations of the dependence of $R$ in $A$ and $\Theta$. (b,c) Comparison between the simulation and experimental results along the horizontal and vertical dashed lines in panel a. (d) A stability test trace over 2 hours.

Simulations and experimental results showed that a simple sine wave modulation of the optical phase was not robust against slowly drifting phase variations (Fig. S3).

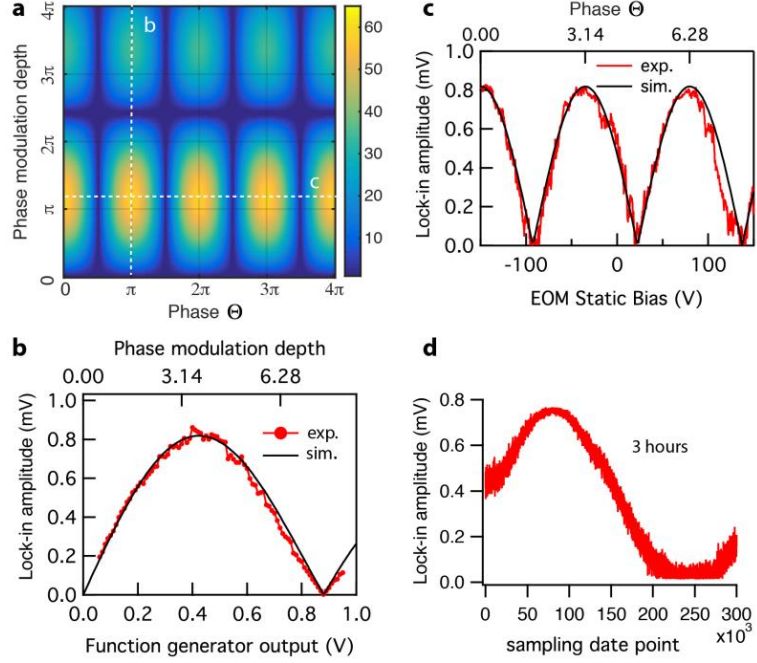

**Figure S3.** Comparison between simulations and experimental results for sine wave phase modulation. (a) Contour plot of simulations of the dependence of $R$ in $A$ and $\Theta$ with the sine wave phase modulation. (b,c) Comparison between the simulation and experimental results along the horizontal and vertical dashed lines in panel a. (d) A stability test trace over 3 hours.

Experiments were performed using a potassium dideuterium phosphate (KD*P) phase modulator with crystal dimensions of 3.5 × 3.5 × 160 mm (M350-160 phase, Conoptics Inc). The EOM driver was Model 275 from Conoptics Inc, which amplified the output of the function generator (4063 BK Precision) by 275 times. KD*P has electrooptic coefficients $r_{63}$ of 26.4 pm/V.[3] The phase change is calculated by $\Delta\phi = \pi l E r_{63} n_o^3 / \lambda$.[4]

S5

**Supplementary Note 2: Nonlinear wave mixing in MoS₂**

Monolayer MoS2 has $D_{3h}$ symmetry and its second-order nonlinear susceptibility tensor has nonzero elements of $\chi^{(2)}_{yyy} = -\chi^{(2)}_{yxx} = -\chi^{(2)}_{xxy} = -\chi^{(2)}_{xyx} = \chi^{(2)}$, where the crystalline coordinate is illustrated in Fig. S4a.[1,5,6] For conventional polarized SHG, the generated second harmonic field perpendicular (parallel) to the incident fundamental field is $E^{2\omega}_{\perp} \propto -\chi E^{\omega} \cos 3\theta$ ($E^{2\omega}_{//} \propto \chi E^{\omega} \sin 3\theta$). The intensities are $I^{2\omega}_{\perp} \propto (\chi E^{\omega} \cos 3\theta)^2$ ($I^{2\omega}_{//} \propto (\chi E^{\omega} \sin 3\theta)^2$). Fig. S4b shows the experimental data in red circles and calculations in black solid line.

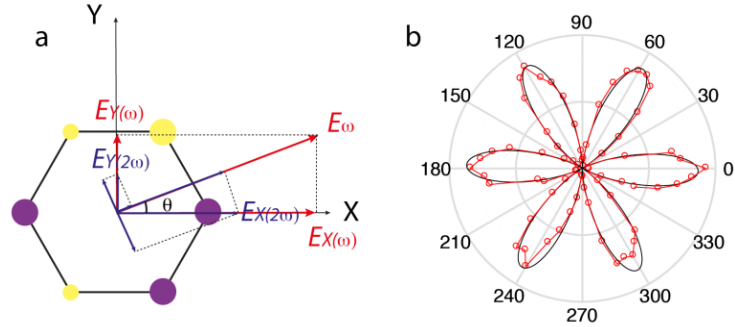

**Figure S4.** Polarized SHG in monolayer MoS₂. (a) Coordinate for MoS₂ crystal and laser fields of the fundamental (in red) and the generated second harmonic filed (in blue). (b) The intensity of the second harmonic in the direction perpendicular to the incident fundamental as function of the angle $\theta$; red circles are the experimental data points, and the black curve is from calculations.

For the case of degenerate parametric amplification, both the fundamental and second harmonic fields are incident. Taking the same symmetry information as above, the field amplitude and intensity changes can be calculated as:

$$\begin{aligned}
\Delta E^{\omega}_y &= i\left(\chi E^{\omega} E^{2\omega}(2\sin\theta\cos\theta)\right)\exp(i\phi^{2\omega} - i\phi^{\omega}), \\
\Delta E^{\omega}_x &= i\left(\chi E^{\omega} E^{2\omega}(\sin^2\theta - \cos^2\theta)\right)\exp(i\phi^{2\omega} - i\phi^{\omega}), \\
\Delta I^{\omega} &= \text{Re}\left(2E^{\omega}_x \Delta E^{\omega}_x + 2E^{\omega}_y \Delta E^{\omega}_y\right)\left(i\exp(2i\phi^{\omega} - i\phi^{2\omega}) - i\exp(-(2i\phi^{\omega} - i\phi^{2\omega}))\right), \\
&= \left(-4\chi E^{\omega} E^{\omega} E^{2\omega} \cos(3\theta)\right)\sin(2\phi^{\omega} - \phi^{2\omega}),
\end{aligned} \quad (S7)$$

and

$$\Delta E_y^{2\omega} = i\left(\chi E^\omega E^\omega(-\cos^2\theta + \sin^2\theta)\right)\exp(2i\phi^\omega),$$

$$\Delta E_x^{2\omega} = i\left(-\chi E^\omega E^\omega(2\sin\theta\cos\theta)\right)\exp(2i\phi^\omega),$$

$$\Delta I^{2\omega} = \mathrm{Re}\left(2E_x^{2\omega}\Delta E_x^{2\omega} + 2E_y^{2\omega}\Delta E_y^{2\omega}\right)\left(i\exp(i\phi^{2\omega} - 2i\phi^\omega) - \exp(-(i\phi^{2\omega} - 2i\phi^\omega))\right),$$

$$= 4\chi E^\omega E^\omega E^{2\omega}\cos(3\theta)\sin(2\phi^\omega - \phi^{2\omega}),$$

(S8)

where $E$ and $\phi$ are the amplitude and phase of the fields labeled with the frequency and the direction. Fig. S5a illustrates the relative directions, and Fig. S5b shows comparison between the experimental results (in solid dots) and calculations (in black line).

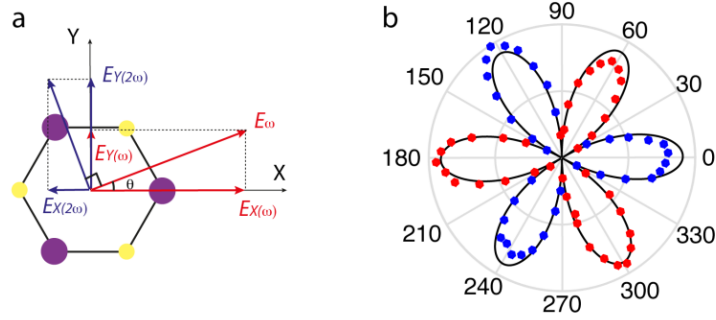

**Figure S5.** (a) The coordinate system for MoS$_2$ crystal, the incident fundamental (in red) and the incident second harmonic field (in blue). (b) Dependence of the lock-in amplitude and phase signals on the orientation of monolayer MoS$_2$ relative to the field polarizations. The red and blue colors indicate a lock-in phase of 0° or 180°, respectively.

**Supplementary Note 3: Determining the absolute MoS$_2$ crystal orientation**

While determination of the relative crystal orientation – including mirror symmetry domains and twins – is easily accomplished by rotating the sample (or polarization) and using a theoretical curve like the one shown in Fig. 2h, determination of the absolute crystal orientation requires an additional step. For instance, Fig. S6 shows two possible correlations between the DPA signal phase (i.e. lock-in phase) and the MoS$_2$ crystal orientation.

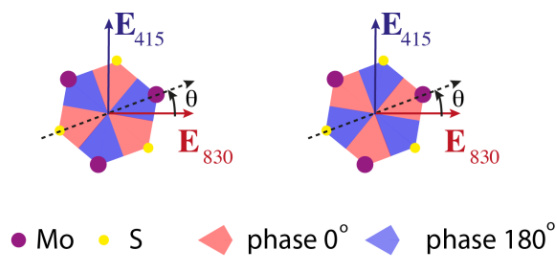

**Figure S6.** Two possible correlations between the DPA signal phase and the MoS$_2$ crystal orientation.

Determination of the absolute crystal orientation can be accomplished by calibrating the lock-in phase using a known standard (such as a BBO crystal). For the case of the multi-grain MoS$_2$ flake shown in Fig. 2c (reproduced here as Fig. S7), we were aided by the additional knowledge that molybdenum terminated edges tend to exhibit sharp straight lines, unlike sulfur terminated edges (compare the edges of grains I and II, for example).[7] Using this information, we determined that a lock-in phase of 180° (blue color) corresponds to $-30° < \theta < 30°$, shown on the left in Fig. S6. We could then assign the crystal orientation of grain I as 57° relative to the polarization direction of the fundamental field (830 nm), and the crystal orientation of the other grains could be determined by their amplitude and phase, with the help of the theoretical curve shown in Fig. 2h. Determining the orientation of grains V, VI, and VII – which are not maximum DPA amplitude domains – was aided by rotating the sample ±5° to assign each grain's unique position on the curve.

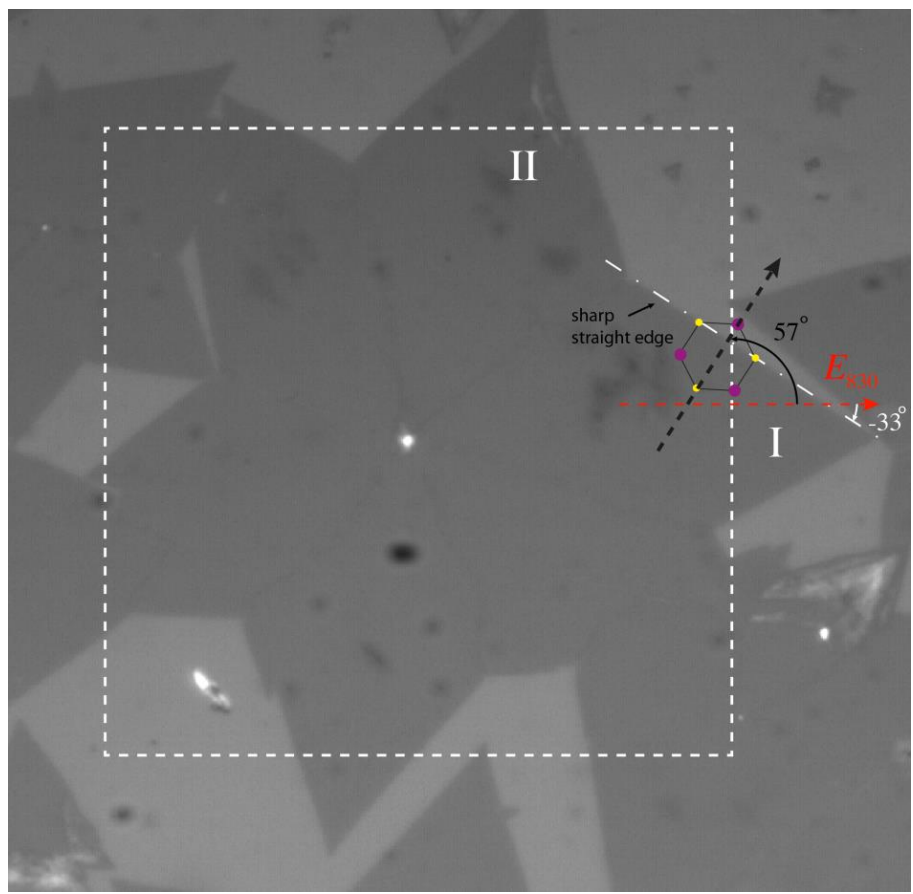

**Figure S7**. Bright-field image of the multi-grain MoS$_2$ flake shown in Fig. 2c-i. The sharp straight edges of the flake are molybdenum terminated, as illustrated here.

**Supplementary Note 4: Spatial resolution of the instrument**

The 830 nm laser was focused to a Gaussian spot with FWHM of 1.1 µm; the 415 nm laser was focused to a Gaussian spot with FWHM of 1.4 µm (Fig. S8). The spot sizes were not diffraction-limited because the laser beams did not fill the back aperture of the objective lens. For DPA, the intensity change is determined by $\Delta I_\omega \propto \chi^{(2)} I_\omega \sqrt{I_{2\omega}}$. Hence, the effective spatial resolution is determined by $I_\omega \sqrt{I_{2\omega}}$, and could be further improved beyond that demonstrated in Fig. 3 of the main text.

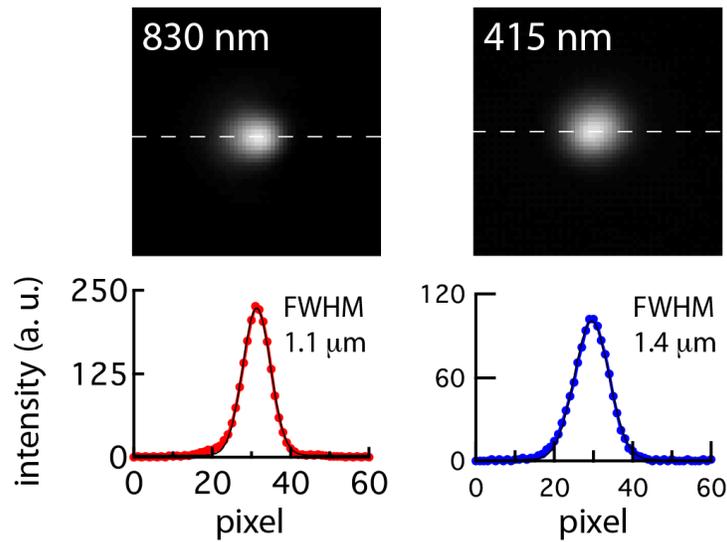

**Figure S8.** Bright-field image of focused laser spots under the microscope objective lens. The black curves are Gaussian fits. The pixel to distance conversion was determined using a stage micrometer (R1L3S2P, Thorlabs).

# Supplementary Note 5: Signal contrast comparison between DPA microscopy and conventional SHG

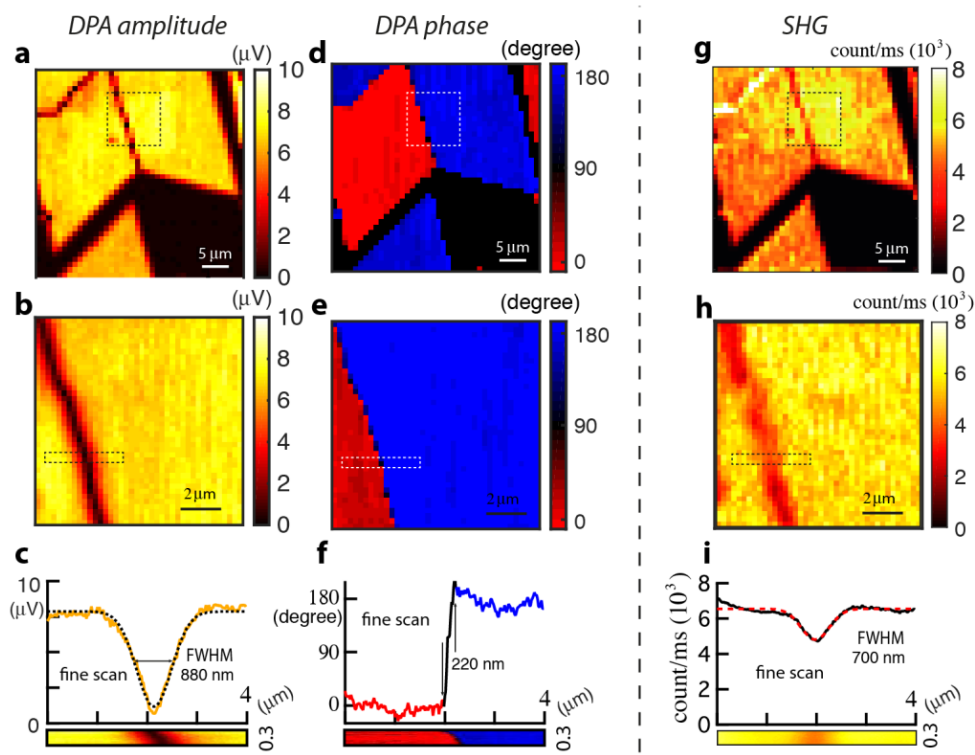

**Figure S9.** Comparison between DPA and SHG mapping of the same MoS$_2$ mirror twin grain boundary and its surrounding area. For both PDA and SHG, the 830 nm laser power was 15 mW, and the dwell time was 1 s for each pixel; for PDA, the 415 nm laser was 0.4 mW.

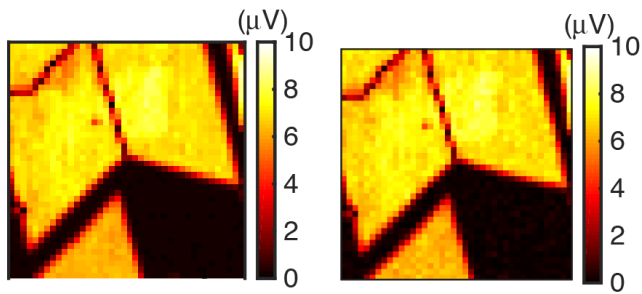

**Figure S10.** Comparison between mapping of the MoS$_2$ mirror twin with pixel dwell time of 1 s (left) and 0.05 s (right). Equivalent image quality suggests an even smaller dwell time can be used to improve scanning speed.

S11

**Supplementary Note 6: Optical and electronic configuration for ω and 2ω detection**

A Ti:sapphire oscillator (Coherent Mira HP) generated 830 nm laser pulses of ~110 fs pulse duration at 76 MHz repetition rate. The $\lambda$ = 830 nm laser beam passed through a small aperture (to produce a near-Gaussian beam), a half-wave plate, a Glan-Taylor polarizer (GT5, Thorlabs), then was focused into a 0.1 mm thick $b$-barium borate crystal (BBO, Type 1, Eksma Optics) by a 50 mm focal length lens (~8 nJ pulse energy) to generate $\lambda$ = 415 nm laser pulses (~40 pJ pulse energy). The output beams were re-collimated by a 50 mm focal length lens, and then separated by a dichroic beam splitter (042-0845，>99.5% reflectivity @ 800nm，Eksma Optics). Residual 830 nm light was removed from the 415 nm beam using a colored glass filter (FGS550, Thorlabs). The 415 nm or the 830 nm beam was sent to the EOM depending on which one was used for detection (see details below and Fig. S11). The beam used for detection bypassed the EOM. One of the two beam went into an optical delay line consisting of mirrors mounted on a linear translation stage (462-X-M, Newport Corporation) driven by a piezoelectric inertia actuator (ZBT225B, Thorlabs). The second harmonic and fundamental paths were re-combined by a beam combiner (042-4805 Eksma Optics, >99.5% reflectivity @ 390-410 nm), and were directed into an inverted optical microscope (Nikon Ti-U). The two beams were focused onto the sample by a microscope objective lens (Nikon, CFI S Plan Fluor ELWD, 40×, 0.6 numerical aperture), and then directed to an amplified Si photo-detector (PDA 36A, Thorlabs) by a 25 mm focus length condenser lens. When the 830 nm (415 nm) laser beam was detected, a long (short) pass filter FGL550S, Thorlabs (FGB 39, Thorlabs) was place in front of the detector. The signal from the photodetector was fed directly into a lock-in amplifier (HF2LI, Zurich Instruments), and the output was recorded and analyzed by a computer. For spatial mapping, the sample was scanned relative to the stationary focal point using a piezo stage (P-545.xR8S PI nano XY Piezo System, Physikinstrumente). For the sample rotation dependent experiments, a motorized precision rotation stage (PRM1Z8, Thorlabs) was used.

A function generator (4063, BK Precision) generated the saw-tooth wave for the EOM driver (Model 275, Conoptics) and supplied the reference signal for the lock-in amplifier. The EOM was a potassium dideuterium phosphate (KD*P) phase modulator (M350-160 phase, Conoptics). To minimize phase noise, the separated beam paths (dual beam paths from separator to combiner, including the EOM) were placed inside a homemade box to minimize the effects of external disturbances. Inside the box, the EOM was placed on a water-cooled breadboard (Thorlabs) fed by a re-circulating chiller (Coherent).

We note that, in principle, it should not matter which beam passes through the EOM; phase modulation of one beam will result in intensity modulation of both beams. However, we obtained better performance when we detected the beam that did not pass through the EOM, as illustrated in Fig. S11. Detecting the beam that passed through the EOM always resulted in higher signal background levels, possibly due to imperfections of the EOM or from intrinsic effects like the piezoelectric effect or the elastooptic effect, which could manifest as very small intensity modulations.

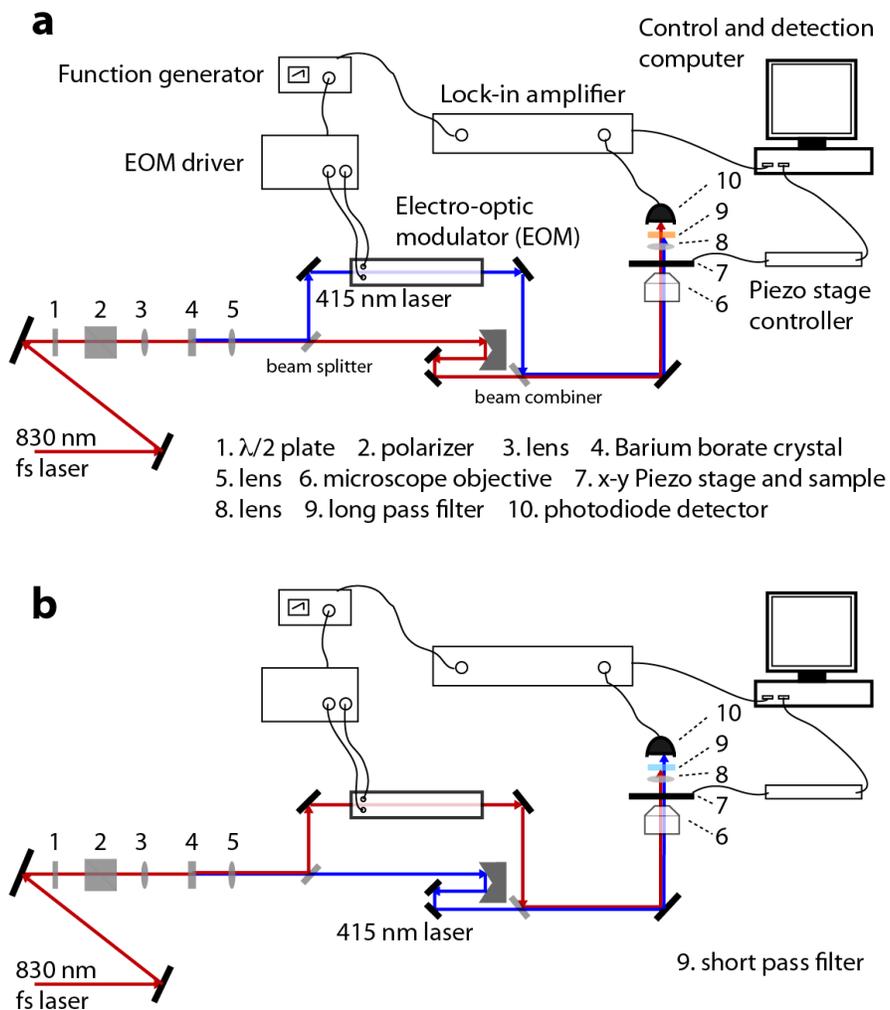

**Figure S11.** Arrangement of optical components and associated electronics for two embodiments of the DPA technique. (a) Detection of the fundamental field ($\lambda = 830$ nm). (b) Detection of the second harmonic field ($\lambda = 415$ nm).

**Supplementary Note 7: Sensitivity of the technique**

In Fig. S12, we compare the signal-to-background ratio for our laboratory implementation of DPA to shot-noise-limited (i.e. background-free) conventional SHG. In our laboratory comparison, the signal-to-noise performance of SHG exceeds that of DPA due to additional electronic noise sources in the DPA measurement. Additionally, we observe a difference in signal-to-noise ratio between 415 nm detection and 830 nm detection. Equation S3 indicates that the amplitude of the lock-in signal should be the same for detection of the fundamental ($\lambda = 830$ nm) and detection of the second harmonic ($\lambda = 415$ nm). The observed difference in the signal-to-noise performance when detecting at 830 nm *vs.* detecting at 415 nm is attributed to the wavelength dependence of the responsivity of the photodiode used for signal detection. The magnitude of the lock-in signal scales linearly with the photodiode responsivity as long as the amplifier is not saturated.

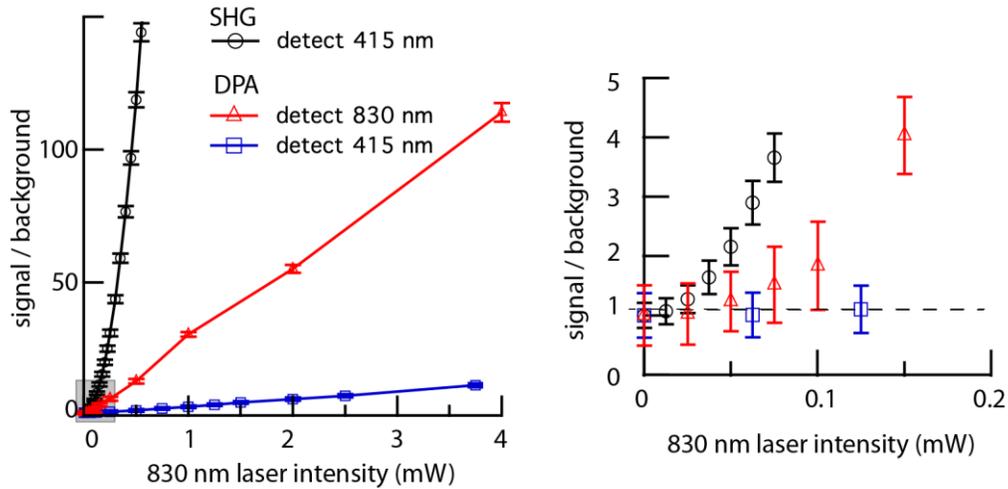

**Figure S12.** Scaling of the signal-to-background ratio for DPA and SHG.

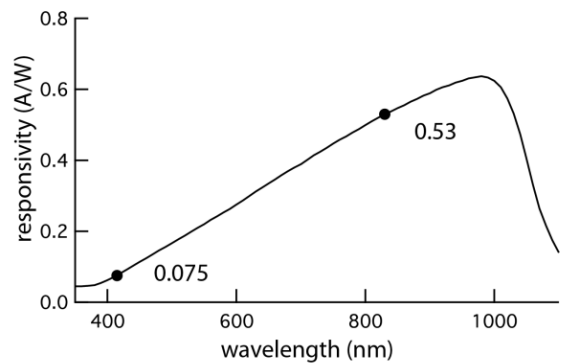

**Figure S13.** Responsivity curve of the PDA 36A photodiode (data from Thorlabs) used in this experiment, with values indicated at the fundamental and second harmonic frequencies.

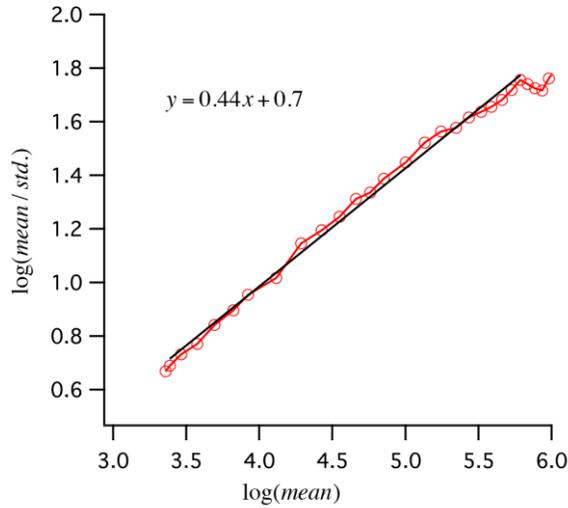

**Figure S14. Shot noise limited optimal SHG detection.** Shot noise limitation is that the signal (mean value) to noise (standard deviation) ratio is equals to square root of the signal (mean value). A linear fit gives a slope of 0.44, close to 0.5. The mean value is the photon counts number from the gated photon counter (SR 400, Stanford Research Systems.).